\documentclass[preprint2]{aastex}
\usepackage{graphicx}

\newcommand{\kms}{km~s$^{-1}$} 
 
\newcommand{\cc}{$^{\circ}$} 
\newcommand{\cds}{cd$^{-1}$\,} 
\newcommand{\cd}{cd$^{-1}$}
\newcommand{\I}{\'\i} 
\newcommand{\DSCT}{$\delta$ Sct}
\newcommand{\dgg}{$\ddagger$}
\shorttitle{Variable stars in the Galactic anticenter direction}
\shortauthors{Poretti et al.}
\begin{document} 
\title{Preparing the COROT space mission: new variable stars in the
galactic Anticenter direction\,$^{\ddagger}$  
}
\author{E.~Poretti\altaffilmark{1}, R.~Alonso\altaffilmark{2}, P.J.~Amado\altaffilmark{3},
J.A.~Belmonte\altaffilmark{2}, R.~Garrido\altaffilmark{3}, S.~Mart\I n-Ruiz\altaffilmark{3,1},
K.~Uytterhoeven\altaffilmark{4}, C.~Catala\altaffilmark{5}, Y.~Lebreton\altaffilmark{6},
E.~Michel\altaffilmark{5},  J.C.~Su\' arez\altaffilmark{3,5}, 
C.~ Aerts\altaffilmark{4}, O.~Creevey\altaffilmark{2,7}, M.J.~Goupil\altaffilmark{5},
L.~Mantegazza\altaffilmark{1}, P.~Mathias\altaffilmark{8}, M.~Rainer\altaffilmark{1}, 
W.W.~Weiss\altaffilmark{9}}
\altaffiltext{1}{INAF-Osservatorio Astronomico di Brera, Via Bianchi 46,
23807 Merate, Italy -- poretti@merate.mi.astro.it} 
\altaffiltext{2}{Instituto de Astrof\I sica de Canarias, C/ V\I a L\'actea s/n, 38200 La 
Laguna, Tenerife, Spain}
\altaffiltext{3}{Instituto de Astrof\I sica de Andaluc\I a, C.S.I.C., Apdo. 3004, 18080 Granada,
Spain}
\altaffiltext{4}{Instituut voor Sterrenkunde, Katholieke Universiteit Leuven, Celestijnenlaan 
200 B, 3001 Leuven, Belgium}
\altaffiltext{5}{Observatoire de Paris, LESIA, UMR 8109, 92195 Meudon, France}
\altaffiltext{6}{Observatoire de Paris, GEPI, UMR 8111, 92195 Meudon, France}
\altaffiltext{7}{High Altitude Observatory, NCAR, Boulder, Colorado, USA}
\altaffiltext{8}{Observatoire de la C\^ote d'Azur, GEMINI, UMR 6203, BP 4229, 06304 Nice Cedex 4, France}
\altaffiltext{9}{Institut f\"ur Astronomie, Universit\"at Wien, T\"urkenschanzstrasse
17, 1180 Wien, Austria} 
\altaffiltext{\dgg}{Based on 
observations collected at the S.~Pedro Mart\I r, Sierra Nevada, Teide,
La Silla, Haute--Provence and Roque de Los Muchachos (Telescopio
Nazionale Galileo and Mercator telescopes) observatories.}
 
\begin{abstract}
The activities related to the preparation of the asteroseismic, photometric space
mission COROT are described. Photoelectric observations, wide--field CCD photometry,
$uvby\beta$ calibrations and further time--series have been obtained at different observatories and
telescopes. They have been planned to complete the COROT programme in the direction of the galactic
Anticenter. In addition to suitable asteroseismic targets covering the different evolutionary 
stages between ZAMS and TAMS, we discovered several other 
variable stars, both pulsating and geometrical. 
We compared results on the incidence of variability in the
galactic Center and Anticenter directions.
Physical parameters have been obtained and evolutionary tracks fitting them have been calculated.
The peculiarities of some individual stars are pointed out.
\end{abstract}

\section{Introduction} 

The definition of the observing programme of the asteroseismic space mission COROT 
(COnvection, ROtation and planetary Transits; Baglin et al. 2002)
requires  a careful evaluation of all the potential targets included in the accessible
field of view. One at a time,
the satellite  will monitor six primary targets   
located in two circles centered at $\alpha=18^{\rm h}50^{\rm m}$,
$\delta=0$\cc\, (i.e., in the direction of the Galactic Center) and
$\alpha=6^{\rm h}50^{\rm m}$, $\delta=0$\cc\, (Anticenter direction).
No more than 10
stars for each 150--day run can be monitored in the Seismo CCDs; therefore,
for each pointing we have one primary target and nine secondary targets. 
Secondary targets have to be found close to the primary ones in a 1.4$^{\circ}$ x 2.8$^{\circ}$ area
and they should 
maximize the coverage of the Hertzsprung--Russell diagram. Indeed, 
to match the scientific profile of the mission, the asteroseismic targets 
(solar--like, $\gamma$ Dor, $\delta$ Sct, $\beta$ Cep, Slow Pulsating B stars,...)
have to be chosen along the Zero--Age Main Sequence (ZAMS). 
In this respect, it is expected that ground--based observations would allow us to 
choose pulsators located in the lower part 
of the instability strip, avoiding the too dense frequency spectrum shown by 
the evolved stars. Ideal COROT candidates should be located 
between the ZAMS and the Terminal Age Main Sequence (TAMS).

Poretti et al. (2003; hereafter Paper~I) described how primary
targets were searched in the Center direction.
In the Anticenter direction the
request was to find suitable secondary targets around already fixed primary ones
in such a way as to minimize the impact on the whole program.
Regarding the specific goal to map the lower part of the instability strip, 
we describe here how we matched it by using a limited number of targets.
We also detected several new variable stars in the COROT
fields.

\section{Observations and data reduction} 

The Anticenter direction was monitored with four different instruments
on different occasions. The
irst photoelectric results were obtained for a preliminary list of
candidate secondary targets using the Mercator telescope
in 2002 February. Three to five measurements per night
were obtained on several targets and some  variables 
could be proposed. Dedicated 
CCD observations obtained with the STARE telescope (Brown et al., in
preparation) in
2002 March allowed us to sharpen our approach.  Covering a
wider area around the targets (6.1\cc\,x\,6.1\cc), 
the STARE monitoring (typically one night for each field) 
allowed us to identify
many suspected variable stars. Therefore, these suspected variables became the 
targets of future observations.
The main differences from the data reduction pipeline described in Brown et al. 
are the use of a $V$ filter and different exposure times. These were
optimized for bright stars  and 
accurate photometry is available for stars brighter than $V$=11.0. 
The original time sampling of the STARE images was less than one
minute; therefore seven consecutive images were averaged to give more accurate
mean magnitudes. The differential magnitudes were transformed into an instrumental $V$
system.
Table~\ref{stare} is the inventory of the new wariables we found. 
\placetable{stare}

On the basis of the results obtained by the analysis of the STARE data,
new photoelectric measurements  were carried out at S.~Pedro Mart\I r 
(SPM: 2002 December 3--10
and 2003 November 14--23) and at Sierra Nevada Observatory
(OSN; on December 20, 2002 and  February 16, 2003).
In both observatories a simultaneous $uvby$ photometer was used.
These observations were performed to confirm the variability of selected
targets on a longer time baseline, to clarify doubtful cases and to 
characterize a little more the pulsational behaviour of well--established ones. 
The reduction of the photometric data and their transformation into the 
standard system were done following the procedures described in 
Olsen (1993) and references therein. The results of these procedures 
applied to our dataset will be presented in a future work (Amado et al., 
in preparation).

To complete the target characterization,
the data from the extensive ground--based survey
carried out at Sierra Nevada Observatory ($uvby\beta$ 
photometry of all stars brighter than $V$=8.0) were used
to build a colour--magnitude diagram (CMD), as we did in
Paper~I. In a second step, we also considered spectroscopic observations,
obtained at the Haute--Provence Observatory (ELODIE and AURELIE instruments at the 
193--cm and 152--cm telescopes, respectively),  at La Silla Observatory (FEROS instrument at
the 152--cm telescope) and at the Italian Telescopio Nazionale Galileo (SARG
instrument). All these data were used to build up the GAUDI
archive (Solano et al. 2005). Line profile variations and double--lines
were searched in the high--resolution spectra. Moreover,
$v\sin i$ determinations were performed; resulting 
uncertainties are of the order of a few \kms. 

To complete the coverage of the HR diagram, we also 
searched for COROT targets among $B$--stars using  the Mercator
telescope.
The time sampling of these data is very different than that of $\delta$ Sct data, 
as the data were taken in the general framework of a long--term monitoring program
with a time spread of 2 years (Aerts et al., in preparation). This type of monitoring is much
better suited to discover gravity mode oscillations in B stars (see, e.g., Aerts
et al. 1999). 
\section{The identification of potential targets} 
\begin{figure}[]
%\resizebox{\hsize}{!}{\includegraphics{/gamma/poretti/alfa/corot/data/fig1/anticmd.ps}}
\plotone{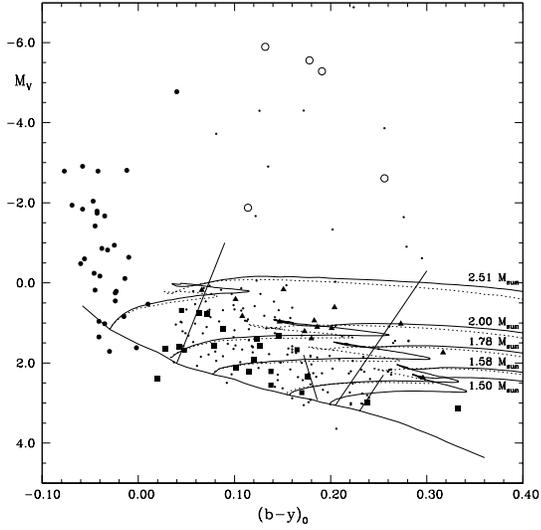}
%\plotone{/gamma/poretti/alfa/corot/data/fig1/anticmd.ps}
\caption{Absolute magnitude $M_V$ vs. $(b-y)_0$
colour index for stars located in the Anticenter direction.
Dotted and solid lines indicate evolutionary tracks for
$d_{\rm over}$=0.1 and $d_{\rm over}$=0.2, respectively.
Solid squares represent stars certainly unevolved, independently of
overshooting influence.  Solid triangles represent stars
whose evolutionary status depends on the importance of overshooting.
Open circles represent stars too advanced on their evolutionary
tracks. Large filled circles on the left indicate blue stars. Small
filled circles  indicate
stars located in the instability strip but too far from the
primary targets and therefore
 not investigated for photometric variability. The borders of the
$\delta$ Sct (longer lines) and $\gamma$ Dor (shorter ones) 
instability strips are also indicated.
} \label{cmd}
\end{figure}

By considering stars belonging to the B, A and F spectral types,
the merging of the $uvby\beta$ OSN survey and STARE photometry
resulted in an initial sample composed of 223 objects. 
Figure~\ref{cmd} shows the CMD as obtained from
$uvby\beta$ photometry performed at OSN.
The same procedures described in Paper~I are used to deredden
colour indices and apparent magnitudes and to determine absolute
magnitudes. ZAMS, $\delta$ Sct 
instability strip borders, evolutionary tracks and models are taken
as in Paper~I and the same symbols as in Fig.~1 of Paper~I have been used;
the borders of the instability strip for $\gamma$ Dor variables are those
reported by Handler \& Shobbrook (2002). 
\placefigure{cmd}

Taking into account the list of potential primary targets in the Anticenter
direction, 150 stars were not considered for further investigations
as they are located too far from the possible satellite pointings. This
decision 
implied that we were obliged to neglect the variables we found among them. 
They are listed in Tab.~\ref{stare} as ``Stars
too far to be secondary targets". Some of them can be considered for further
studies. For example, the ELODIE spectrum of HD 44333 shows that the
star is  a spectroscopic binary; moreover, an eclipse with a minimum light at 
Hel.~JD~2452963.92 has been observed at SPM. In addition 
HD 292962 is noted as a double or multiple star in SIMBAD: the
star can be used to investigate the connection between pulsation
and duplicity.

\begin{figure}[]
% LIMITI FILE PS   18 170 592 718
%\resizebox{\hsize}{!}{\includegraphics{/gamma/poretti/alfa/corot/paper/HD43587out.ps}}
\resizebox{\hsize}{!}{\includegraphics{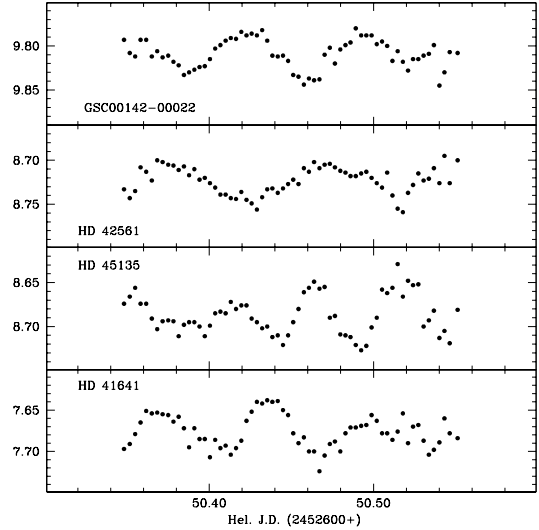}}
\caption{Light curves of some new $\delta$ Sct stars discovered by STARE in the Anticenter
direction. 
%Unfortunately, none of them can be a secondary target: HD 41641, HD 42561 and
%HD 45135 are too far from primary targets, GSC 00142-00022 is too faint to give
%the requested S/N. 
Magnitude scale is the STARE $V$--instrumental system.
}
\label{fdue}
\end{figure}

The remaining 73 stars fall close to the primary targets.
However, some of them  are too faint to be monitored in
the COROT Seismo CCDs: a magnitude fainter than $V$=9.5 will result in a low S/N,
precluding the possibility of doing asteroseismology at the $\mu$mag level.
They are listed in Tab.~\ref{stare} as ``Stars
too faint to be secondary targets".
We just note that GSC 00144-03031 is 
a double--mode pulsator 
(Poretti et al., in preparation). 
Figure~\ref{fdue} shows some examples of light curves of pulsating stars discovered
in our survey that cannot 
be included in the list of secondary targets as they are too faint and/or 
too far from primary ones. 
The variable stars that fit the
requirements about brightness and distance from the primary targets 
are listed in Tab.~\ref{stare} as ``Potential secondary targets".
\placefigure{fdue}

\begin{figure}[]
%\resizebox{\hsize}{!}{\includegraphics{/gamma/poretti/alfa/corot/paper/e3.ps}}
\resizebox{\hsize}{!}{\includegraphics{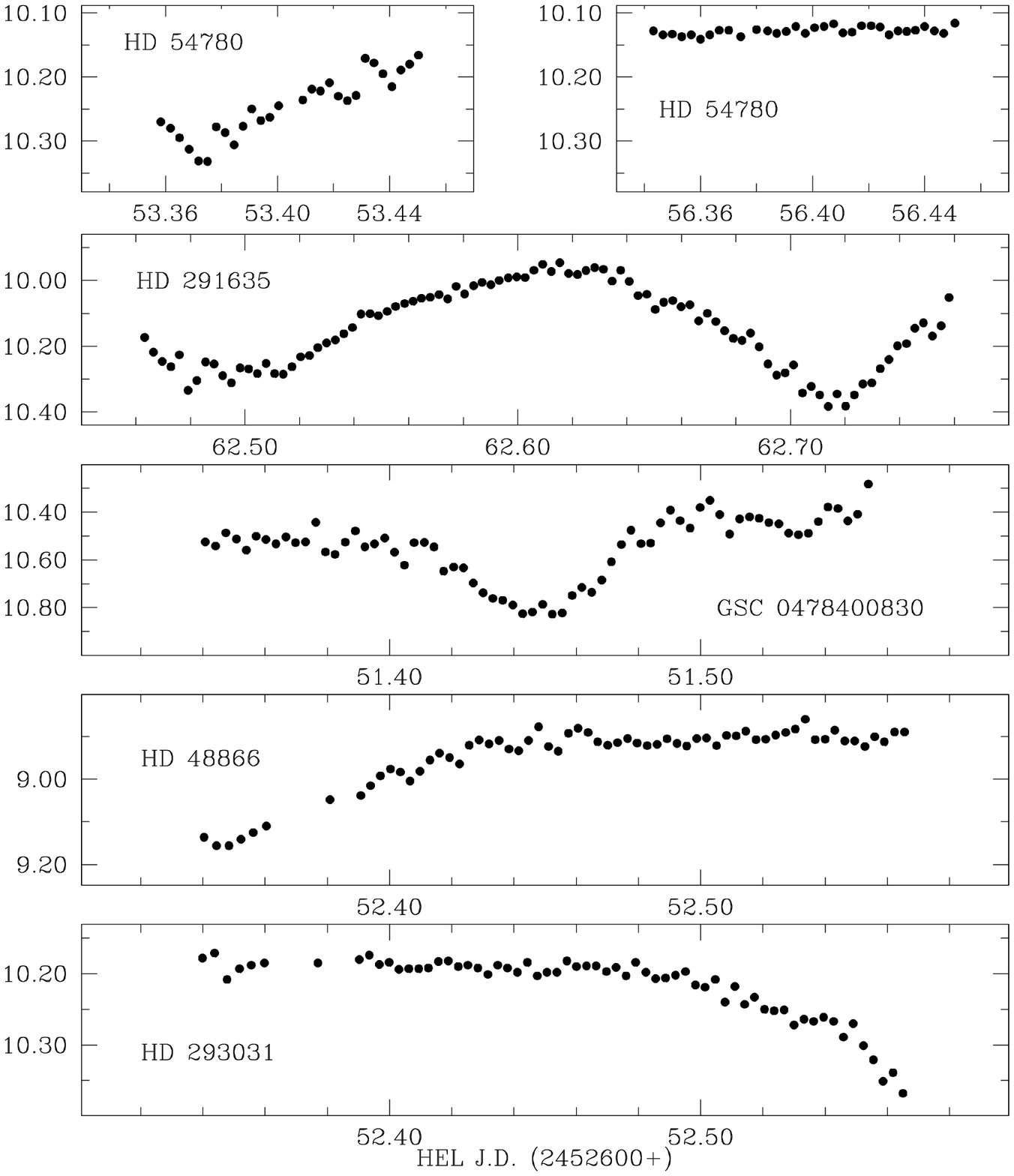}}
\caption{Light curves of some new eclipsing binaries  discovered by STARE.
Magnitude scale is the STARE $V$--instrumental system.}
\label{eb}
\end{figure}

We stress the fact that all doubtful cases have been omitted, to avoid producing
false alarms.
This means that small amplitude pulsating stars with a poorly defined
light curve as well as geometrical variables simply showing a drift are not
considered, as both effects can be due to random and systematic effects in
the wide--field STARE photometry. On the other hand, long period and red variables are stars
showing a well-defined nightly drift or different mean magnitudes 
(one field was  monitored on two nights separated by 2~d). 
To illustrate a few other examples of variables, Fig.~\ref{eb} shows light curves
that strongly support the eclipsing binary  hypothesis. Of course, no 
period can be given as most of the stars
were monitored on one night only.
The whole STARE photometry,  as well as the $uvby\beta$ one, 
will be available in the GAUDI archive (Solano et al. 2005).
\placefigure{eb}

\begin{deluxetable} { ll l r l}
\tabletypesize{\footnotesize}
\tablewidth{0pt}
\tablecaption{New variable stars discovered  
in the Anticenter direction.\label{stare} }
\tablehead{
\colhead{Star} & 
\colhead{$V$\tablenotemark{a}} & \colhead{Sp.}  
& \colhead{Ampl.\tablenotemark{b}}
& \colhead{Type\tablenotemark{c}} \\
\colhead{} & 
\colhead{} & 
\colhead{} & 
\colhead{[mmag]} & \colhead{and remarks}  }
\startdata
\multicolumn{5}{c}{Potential secondary targets} \\
HD ~\,43021  & ~\,7.84 & A0 & 45  & \DSCT \\
HD ~\,43286  & ~\,7.02 & B5 & 35  & Geometrical ? \\
HD ~\,44195  & ~\,7.56 & F0 & 30   & \DSCT, $\gamma$ Dor too ? \\
HD ~\,44283  & ~\,9.36 & F5& $<$50    & \DSCT, small $v\sin i$, small frequency range\\ 
HD ~\,44562  & ~\,8.63 & A3 & 20    & \DSCT \\
HD ~\,44872  & ~\,8.40 & A3 & 20    & \DSCT \\
HD ~\,45196  & ~\,8.33 &    & 14& Geom., fast rotator\\
HD ~\,48719  & ~\,9.19 & A5 & 30  & \DSCT, evolved but fast rotator \\
HD ~\,50844  & ~\,9.10 & A2 & 80  & \DSCT, highest amplitude \\
HD ~\,50870  & ~\,8.88 & F0 & 70  & \DSCT \\
HD ~\,55113  & ~\,8.70  & K5 & $\sim$20   & Red variable \\
HD 291684 & ~\,9.83 & A0 & 60 &  \DSCT, ZAMS object \\ 
HD 293340 & ~\,9.53 & F0 & 40  & \DSCT \\
\multicolumn{5}{c}{Too faint to be secondary targets} \\
HD ~\,54780  & 10.17 & A0    &  $\sim$220 & E \\
HD 289732 & 10.8 & B8 & 105 & $\beta$ Cep ? multiperiodic\\
HD 291635 & 10.4 & B    & 400 & E \\
HD 291791 & 10.5 & F0 & $\sim$80 & \DSCT\\
HD 292402 & 10.10 & F0 &120 & \DSCT\\
HD 292525 & 10.36 & F2 &  150& \DSCT\\
HD 292864 & 10.45 & A2   &  75 & \DSCT \\
HD 292930 & 10.33 & F0V&  40 & \DSCT\\
HD 292971 & 10.12 & F2V    &  100& \DSCT\\
GSC 00142-00022 & 10.3 &   &65 & \DSCT \\ 
GSC 00143-01718 & 10.3     &    & $\sim$200  & HADS or E  \\
GSC 00144-03031 & 10.1 &   &$>$400   &HADS, double--mode\\ 
GSC 00143-00139 & 10.8 &  & $\sim$120& HADS or E \\
GSC 04784-00830 & 11.5 &      & 600 & E \\
GSC 04814-00028 & 11.1 &   &$\sim$100  & Red variable\\
\multicolumn{5}{c}{Too far to be secondary targets} \\
HD ~\,41641 &~\,7.86 & A5 & 70 & \DSCT, SB2 ? \\ 
HD ~\,42561 &~\,8.89 & A2 & 60 & \DSCT\\
HD ~\,44333 &~\,6.31 & A4.5V &  & E+SB2 \\
HD ~\,45135 &~\,8.82 & A2 &  100& \DSCT\\ 
HD ~\,48866 & ~\,9.0  &  A0   &  $\sim$300 & E  \\
HD ~\,52239 & ~\,9.06 & A5 & 55 & \DSCT\\
HD ~\,54331  & ~\,7.94  & A0   &$\sim$40  & Long period\\
HD 292962 & ~\,9.89 & F0V  & 200   & \DSCT \\
HD 293031 & 10.48~ &  F8III   & $>$200  & E\\
HD 293622 & ~\,9.98 & A & 35 & \DSCT\\
\enddata
\tablenotetext{a}{Values reported in the SIMBAD and/or GAUDI databases; in particular for
GSC stars see GSC version 1.2}
\tablenotetext{b}{Peak--to--peak amplitudes are in the $V$ STARE 
instrumental system; for HD 44195 it is in $y$--light, for
HD 43286 it is in the $V$--light of the Geneva system.} 
\tablenotetext{c}{
HADS stands for High Amplitude $\delta$ Sct star,
E for eclipsing binary} 
\end{deluxetable}
\section{The characterization of the new variables}
\begin{figure}[]
%\resizebox{\hsize}{!}{\includegraphics{/gamma/poretti/alfa/corot/data/fig1/summary5.ps}}
\resizebox{\hsize}{!}{\includegraphics{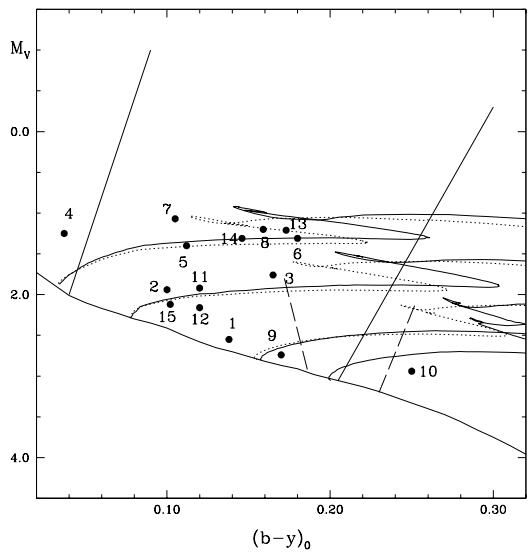}}
\caption{
The $\delta$ Sct variables (1 to 8) and  the suspected $\gamma$ Dor variables
(9 and 10) accepted as targets in the
COROT program,  shown in the CMD. The new pulsating stars discovered in the 
Anticenter direction (11 to 15) are also shown.}
\label{summary}
\end{figure}

As a further step, the rapid variability of the most interesting cases 
was investigated at OSN and SPM sites in dedicated observing runs; among
these stars,
none was previously known as variable and therefore their characterizations
are interesting on their own, independently from their use as COROT targets.
Unfortunately, for some stars (HD 44562, HD 44872, HD 54277,
HD 293340, HD 50870, HD 55113 and GSC 00143-01718) 
the detection of variability on the STARE frames remains the only (but well proven)
 evidence.
Both STARE and Mercator photometry support the variability of HD 43021.
The time series collected on other stars are more
numerous and 
they allowed us a more complete characterization of their variability. The
time series have been analysed by using the least--squares power spectrum method
(Vani\^cek 1971). Moreover, $uvby\beta$ photometry has been extended to
$\delta$ Sct stars fainter than $V$=8.0, either
with dedicated observations at OSN or after the Hauck \& Mermilliod (1998) catalog.
The physical parameters for the new $\delta$ Sct stars in the Anticenter direction 
(Tab.~\ref{dsct}; stars in the upper part plus HD 44195, HD 44283 and HD 50870)
have been derived from $uvby\beta$ photometry only
({\sc templogg} method, Rogers 1995; see also Kupka \& Bruntt 2001),
disregarding for the moment other methods such as $M_V$ determinations from {\sc hipparcos} 
parallaxes or T$_{\rm eff}$ values from spectroscopy. 
Uncertainties on the parameters derived from Str\"omgren photometry are
$\pm$200 K, $\pm$0.2 dex and $\pm$0.2 dex on T$_{\rm eff}$, $\log g$ and [Fe/H], respectively.
Our high--resolution spectroscopy
shows that HD 41641 could be a double--lined spectroscopic binary and therefore
the physical parameters in Tab.~\ref{dsct} are uncertain.
Figure~\ref{summary} shows the position of the new pulsating stars in the CMD.
\placefigure{summary}
\subsection{HD 50844}
\begin{figure}[]
%\resizebox{\hsize}{!}{\includegraphics{/gamma/poretti/alfa/corot/paper/cand5.ps}}
\resizebox{\hsize}{!}{\includegraphics{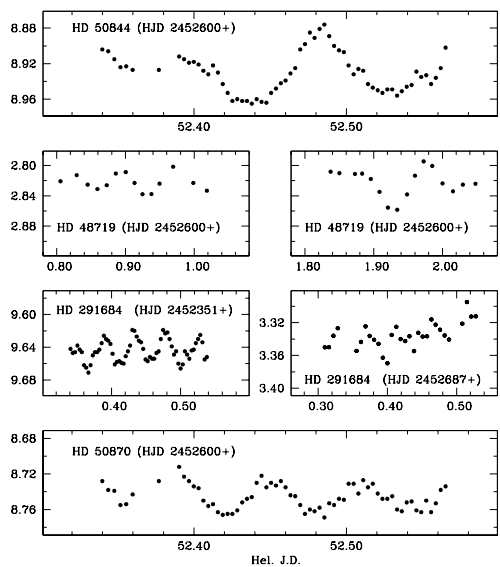}}
\caption{Light curves of some new $\delta$ Sct stars 
that are potential COROT targets. The magnitude scale of
HD 50844, HD 291684 (JD 2452351) and
HD 50870 is the STARE $V$--instrumental system; magnitude
scale of HD 48719 and
HD  291684 (JD 2452687) is differential $\Delta v$ photometry.}
\label{fqua}
\end{figure}

\placefigure{fqua}
High--resolution spectroscopy reveals that
the line profiles are very perturbed: the star
is a multiple one or, more probably, a $\delta$ Sct showing high--degree
modes.  The considerable photometric amplitude (Fig.~\ref{fqua}, top panel)
also suggests the presence of a dominant low--degree mode. It is a  
2M$_{\sun}$, slightly evolved object (star 14 in  Fig.~\ref{summary}).

\subsection{HD 44283}
\begin{figure}[]
% LIMITI FILE PS   18 144 450 650
%\resizebox{\hsize}{!}{\includegraphics{/gamma/poretti/alfa/corot/paper/HD44283.ps}}
\resizebox{\hsize}{!}{\includegraphics{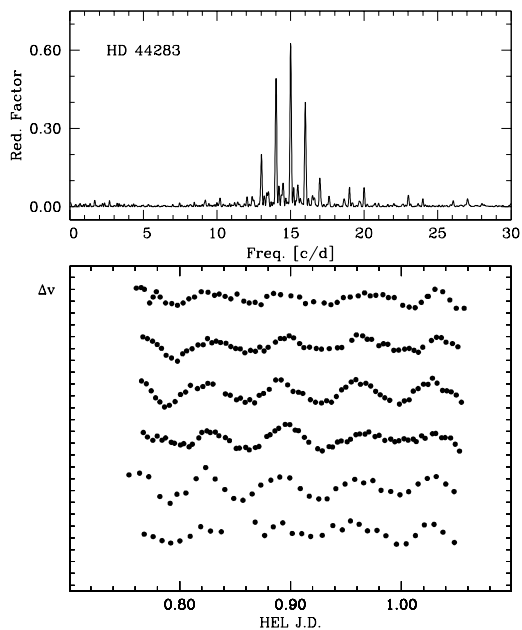}}
\caption{Power spectrum and light curves of the new $\delta$ Sct variable HD 44283.
Light curves were obtained at SPM on JD 2452960, 2452961, 2452962,
2452964, 2452966 and 2452967 nights (from top);
ticksize on the $\Delta v$--magnitudes axis is 0.02 mag.}
\label{multi}
\end{figure}

The intensive monitoring carried out in November 2003 at OSN and SPM allows us
to detect a dominant peak at $f_1$=15.00~\cd (Fig.~\ref{multi}). Other peaks in the 14--16 \cds interval
are noted after introducing $f_1$ as known constituent. No other
term at the highest or lowest frequencies is detected, thus the excited modes are
confined in a well--defined region. The $v\sin i$ value is quite low (19~\kms)
and the star is located in the middle of the instability strip, in a position
compatible both with an evolved and an unevolved status (star 6 in  Fig.~\ref{summary}); 
the frequency regime
is more compatible with the latter hypothesis. Erroneously reported
as a K0 star in SIMBAD, it is actually an F5 star. 
\placefigure{multi}
%{\bf \subsection{HD 43317}
%The few data collected with the Mercator telescope are clearly revealing that
%the star is a new $\beta$ Cep variable.  In particular, the  $U-B$ colour curve
%is in phase with the $V$ one, as expected for these variable stars. 
%The only detectable frequency is 5.15~\cd; the star shows a moderate $v\sin i$
%value, i.e., 130~\kms. }
\subsection{HD 45196}
The frequency analysis of the OSN and SPM data shows a power spectrum where
the signal is confined in the $f\le 6$~\cds region; small amplitude, long
period fluctuations are detectable in the light curves.
The star is a very fast rotator ($v\sin i$=200~\kms; ELODIE and AURELIE
spectra).
It is probably a geometrical
variable, maybe an ellipsoidal one, as also suggested by the frequency
analysis of the $b-y$ colour index.
\subsection{HD 291684}
Both STARE and OSN data show a well--defined light variability (Fig.~\ref{fqua}).
 The star
is located very close to the ZAMS (star 15  in  Fig.~\ref{summary}).
 A rapid, well--defined variability (0.04--0.05~d) 
can be inferred from the STARE night, but the OSN night shows the action
of other terms changing the light curve shape from one cycle to the next. 
\subsection{HD 48719}
Its position in the CMD diagram is 
close to the TAMS and 
superposed on the zigzags of the evolutionary tracks of  2M$_{\sun}$ models
(star 13 in  Fig.~\ref{summary}).
The detected frequency is compatible with an evolved stage ($f_1$=10~\cd),
but the rotational velocity is still fast ($v\sin i$=197~\kms).
The star also seems to be a multiperiodic one (Fig.~\ref{fqua}).
\subsection{HD 44195}
\begin{figure}[]
%\resizebox{\hsize}{!}{\includegraphics{/gamma/poretti/alfa/corot/paper/HD44195rev.ps}}
% LIMITI FILE PS   18 144 450 650
\resizebox{\hsize}{!}{\includegraphics{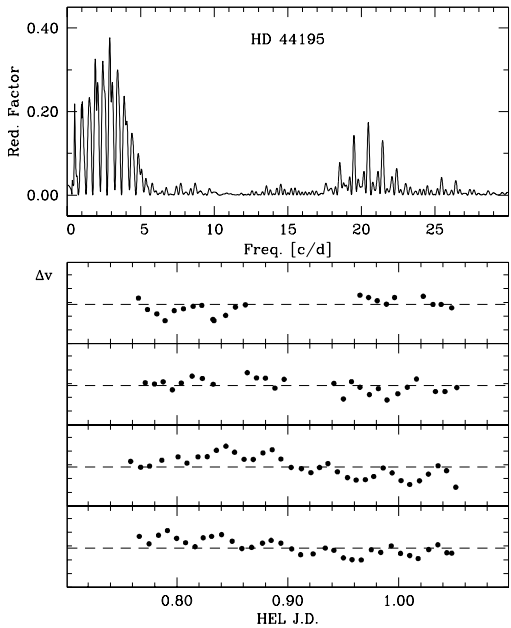}}
\caption{Power spectrum and light curves of the new $\delta$ Sct variable HD 44195.
Peaks at low frequencies also suggest a $\gamma$ Dor pulsation. Measurements were obtained
at SPM on JD 2452963,
2452964, 2452966 and 2452967 nights (from top); ticksize on the $\Delta v$--magnitudes
axis is 0.01 mag. A dotted line in each panel
indicates the mean magnitude of the whole dataset.}
\label{both}
\end{figure}

The power spectrum of the SPM data is quite peculiar, as it shows peaks
at low frequencies and a well--defined peak at around 20~\cds (upper panel
in Fig.~\ref{both}).
The $v\sin i$ value is moderate (58~\kms) and the star is located in the
middle of the instability strip, very close to the ZAMS
(star 1 in  Fig.~\ref{summary}). It is a good
candidate to be a combined $\gamma$ Dor and $\delta$ Sct variable. The ELODIE
high--resolution spectra confirm the presence of bumps all along
the lines, which was already suspected from the AURELIE spectra (Mathias et al. 2004).
It should be noted that the star's light variability has been investigated at SPM 
to confirm the spectroscopic one, since the STARE light curve did not give
a definitive result (the full amplitude of the rapid pulsation is less 
than 0.01 mag). 
The lower panel in  Fig.~\ref{both} shows the light
curves: the mean magnitude is also indicated to evidence the night--to--night
variations originating the peaks at low frequencies in the power spectrum.
\placefigure{both}
\section{Probing the Anticenter direction using $\delta$ Sct stars}
\begin{figure}
\resizebox{\hsize}{!}{\includegraphics{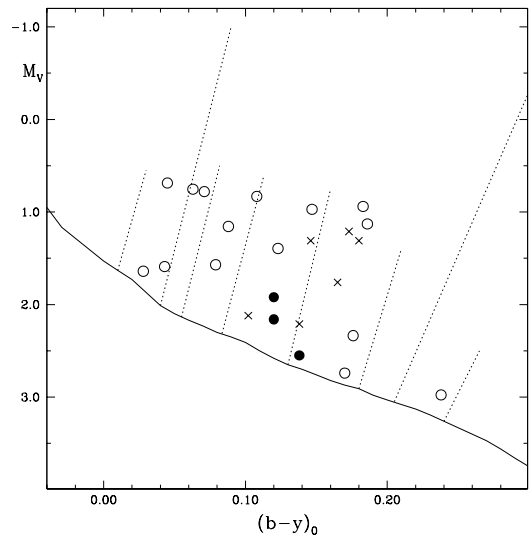}}
\caption{Incidence of $\delta$ Sct variability in the Anticenter sample. The spacings are selected as
in Paper~I, i.e.,
taking the borders roughly parallel to the blue and red borders (the
longest ones) of the instability strip. Filled circles: variable stars brighter
than $V$=8.0. Crosses: variable stars fainter than $V$=8.0.
Open circles: constant stars.}
\label{varwin}
\end{figure}

The sample in the Anticenter direction is 
homogenous up to $V=8.0$, i.e., the limit for which the $uvby\beta$
sample is complete. Therefore, we cannot  evaluate
the incidence of variability since we have only  3 variables and
15 constant stars brighter than $V=8.0$ (Fig.~\ref{varwin}).

On the other hand, we can study  
the positions of $\delta$ Sct stars inside the instability strip considering also 
fainter stars 
since we performed $uvby\beta$ photometry for variable stars. 
They are concentrated in the central part,
in the narrow range 0.10$<(b-y)_0<$0.16. This confirms the result
described in Paper~I: we have  a higher probability to
find $\delta$ Sct variables in the middle of the instability strip rather than close
to the borders. In the Anticenter direction
the $\delta$ Sct stars (Fig.~\ref{varwin}, filled circles and 
crosses) have a distribution similar to the one   
in the whole Galaxy (see Fig.~5 in Paper I).  Moreover,
no variable has been found close or outside the blue border, the
two cases found in the Center survey (HD 170782 and HD 183324; Paper~I) still remain isolated. However, we
note that very--short period $\delta$ Sct stars should populate that region; 
recently  Amado et al. (2004)
discovered an 18--min pulsation in HD~34282, finding evidence that high--radial order pulsators
can be located close to/outside the blue border. Moreover, we remind the reader that
we found an overabundance of variables in the half toward the blue border when
observing in the Center direction (see Fig.~4 in Paper~I); not being found here,
that overbundance still seems to be a peculiarity of the solar neighbourhood in the Center direction. 
COROT photometry from Exoplanetary CCDs, which
will be obtained in both directions within the same magnitude limits,
should supply a more consistent statistics to evaluate this different
abundance. 
\placefigure{warwin}

The CMD shown in Fig~\ref{cmd} is also characterized by a plume of B stars at
$b-y\,<\,0.0$. Most of the time series consist of one night of monitoring
by STARE only. Such a survey should be  able to detect rapid variability,
while slow one should remain
undetectable. Actually, only the faint star HD 289732 shows rapid variability: its
light curve suggests  multiperiodicity.
The distribution of the $v\sin i$ values is quite uniform,
with a preference for slow rotation ($v\sin i <$ 50~\kms). 
\section{Conclusions}
\begin{figure}
\resizebox{\hsize}{!}{\includegraphics{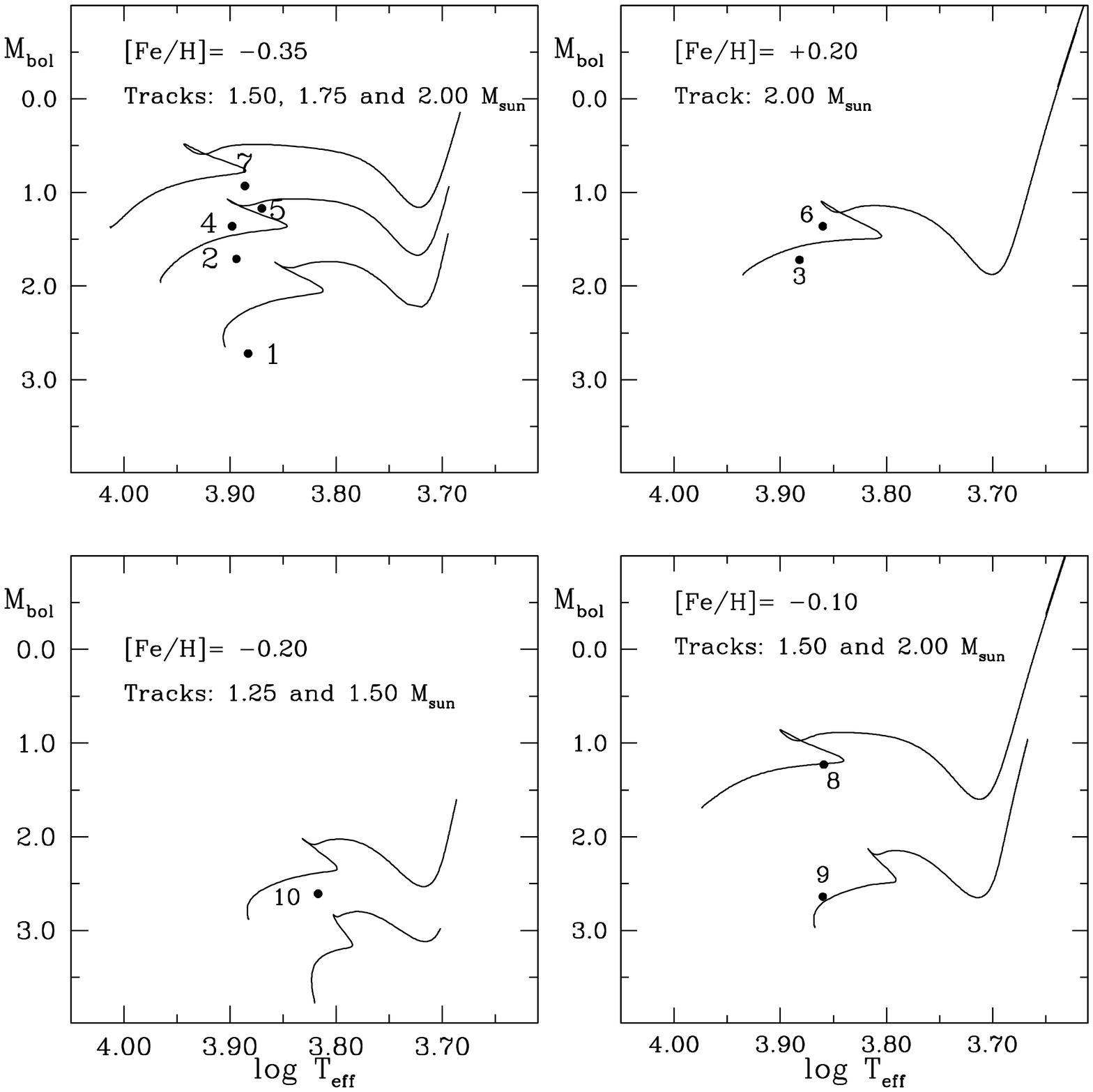}}
\caption{COROT targets in the $T_{\rm eff}-M_{\rm bol}$ plots.
The evolutionary tracks were calculated for the value of [Fe/H] indicated
in the top left corner of each panel. The lower track indicates the model
with the lower mass. For identification of the stars, see Tab.~\ref{dsct}.}
\label{yve}
\end{figure}

\placetable{dsct}
\placefigure{yve}
\begin{deluxetable}{ rl r r r r r}
\tablewidth{0pt}
\tablecaption{Physical parameters of the new pulsating stars. \label{dsct}}
\tablehead{\colhead{Id.\tablenotemark{a}} &
\colhead{Star} & 
\colhead{$v\sin i$} & \colhead{$M_V$\tablenotemark{b}}  
& \colhead{T$_{\rm eff}$}
& \colhead{$\log g$} 
& \colhead{[Fe/H]} \\
\colhead{} & 
\colhead{} & 
\colhead{[\kms]} & \colhead{}  
& \colhead{[K]}
& \colhead{} 
& \colhead{}  }
\startdata
\multicolumn{7}{c}{$\delta$ Sct stars in the Anticenter direction\tablenotemark{c}}\\
11 & HD ~\,41641 & 29 & 1.92 & 7700 & 3.9 & --0.2 \\
12 & HD ~\,43021 & 80 & 2.16 & 7850 & 4.0 & --0.5 \\
13 & HD ~\,48719 & 197&  1.21& 7250 & 3.5 &   0.0 \\
14 &HD ~\,50844 &    &  1.31&  7500 & 3.6& --0.4 \\
15 & HD 291684&    & 2.12&  7600 & 4.0 &  0.0 \\
\noalign{\smallskip}
\multicolumn{7}{c}{$\delta$ Sct and $\gamma$ Dor stars: COROT targets\tablenotemark{d}}\\
1 & HD ~\,44195 & 58 & 2.72 & 7650 & 4.2 & --0.3 \\
6 & HD ~\,44283 & 19 &  1.31&  7250&  3.6 &  0.2 \\
9 & HD ~\,49434& 90 & 2.63 & 7250 & 4.1 & --0.1 \\
3 & HD ~\,50870 & 17 &  1.67&  7600&  3.9&   0.2 \\
4 & HD 170782&198 & 1.35 & 7900 & 3.8 & --0.4 \\
5 & HD 170699&$>$200&1.15& 7400 & 3.5 & --0.4 \\
7 & HD 172189&    &  0.90& 7700 & 3.6 & --0.3 \\
10& HD 171834&72 & 2.64 & 6550 & 4.0 & --0.2 \\
2 & HD 181147&    & 1.70 & 7850 & 3.9 & --0.3 \\
8 & HD 181555 & 170 & 1.20&7200 &3.5 & --0.1  \\
\enddata
\tablenotetext{a}{Identification for Figs.~\ref{summary} and ~\ref{yve}}
\tablenotetext{b}{Absolute magnitudes calculated from $uvby\beta$ photometry}
\tablenotetext{c}{Variables far from COROT primary targets}
\tablenotetext{d}{COROT targets selected both in the Anticenter and Center directions}
\end{deluxetable}

Using the {\sc templogg} method (see Sect.~4), we also calculated 
the physical parameters for targets in the Center direction
(Tab.~\ref{dsct}): the variabilities of HD 181555\footnote{
The position of this star in the CMD
has been revised (see Fig.~8 in Paper~I) on the basis of a more accurate set of $uvby\beta$ photometry.
However,  there is a large  
discrepancy between the Str\"omgren $M_V$ and {\sc hipparcos} parallaxes. They
can be reconciled 
admitting an error of 0.065 mag in the $\beta$ value, which seems huge for such an index. On the
other hand, parallax measurement could be inaccurate owing to two objects which appear
very close to HD 181555 in the Guide Star Catalog.},
 HD 170782 and HD 170699 have been 
reported in Paper~I, while that of 
HD 181147 and HD 172189 have been detected 
when observing these fields. 
The case of HD 49434 is described by Bruntt et al. (2002).
We also note that HD 171834 is photometrically constant from the ground 
(see Tab.~2 in Paper~I), but  
a new, dedicated spectroscopic time series (35 spectra in 4~hours) 
allowed us to detect line profile variations having a relative
amplitude of $3\cdot10^{-4}$; such very
weak line profile variations are also reported by 
Mathias et al. (2004). 
The slight metal--poor content of the variables listed in Tab.~\ref{dsct}
is in agreement with the distribution of metallicities observed in the Solar
neighbourhood (Nordstr\" om et al. 2004).

To give a different picture
of the evolutionary scenario covered using these targets, we compared the physical parameters derived from
$uvby\beta$ photometry with the evolutionary tracks calculated with the CESAM code (4th version,
Morel 1997). 
In Fig.~\ref{yve} the $\log$~T$_{\rm eff}$ and $M_{\rm bol}$ values of each variable star 
are plotted together with the evolutionary tracks calculated for [Fe/H] values similar to the
ones of the variables. An overshooting parameter $d_{\rm over}$=0.2 has been considered and the (small)
bolometric corrections have been introduced (VandenBerg \& Clem 2003). 
All the stars are between ZAMS and TAMS and
most of them have mass greater than 1.50~M$_{\sun}$. 
The presence of some fast rotators among the targets (HD 181555, HD 170782 and
HD 170699) will constitute a severe test for the recent progress
in the treatment of pulsation and fast rotation (Su\'arez et al. 2004).  
The two $\gamma$ Dor variables HD 49434 and 
HD 171834 (stars 9 and 10) are the less massive stars: the latter is more
evolved than the former. 
Their spectroscopic variability, as well as 
the photometric variability
of HD 44195, has some theoretical
implications about the location of the borders of the $\gamma$ Dor instability strip
(see Fig.~7 Handler \& Shobbrook 2002 and Fig.~1 in Kaye et al. 2004).

All the evolutionary stages between ZAMS and TAMS
(both included) are covered (Fig.~\ref{yve}) by considering only stars
brighter than $V$=9.5 and located in two arbitrary directions.
Since this task has been fulfilled using only 10 objects, the other parts of the
ZAMS can be adequately covered by the remaining 83\% of the COROT targets.
This proves that unevolved or slightly evolved pulsating stars 
are quite common and they can be included as 
a solid, ground--based tested baseline in any asteroseismic mission from space. 
We can look at the different targets as a sort of key--stops along the stellar 
evolution path; the
possibility to sound their interiors by detecting oscillations at the $\mu$mag level
in the COROT time series will result in a great improvement in the stellar physics knowledge. 
\acknowledgements 
This research has made use of the {\sc simbad}, VizieR (GSC 1.2) and {\it Aladin} 
databases, operating at 
CDS, Strasbourg, France, and of {\sc gaudi}, the data archive and access system
of the ground--based asteroseismology programme of the {\sc corot} mission. The
{\sc gaudi} system is maintained by {\sc laeff} which is part of the Space
Science Division of {\sc inta}. The authors wish to thank the {\sc stare} team 
for the attribution of observing time to the {\sc corot} project.
The authors wish also to thank 
G.~Handler  for useful comments and 
J.~Vialle for the careful proofreading of the first draft of
the manuscript.
RA and JAB acknowledge financial support from grants {\sc a}y{\sc a2001-1571} and 
{\sc esp2001-4529-pe} of the 
Spanish National Research plan. KU and CA are supported by the Research Fund of
the Katholieke Universiteit Leuven (grant {\sc goa/2003/4}).
SM acknowledges financial support from a European Union Marie Curie
Fellowship, under contract {\sc hpmf-ct-2001-01146}. RG ad JCS 
acknowledge financial support from the programme {\sc esp2001-4528-pe}. 
WW was supported by the Austrian Fonds zur F\"orderung der
Wissenschaftliche Forschung ({\sc p14984}) and the {\sc bm:wuk} (project {\sc corot}).
PJA acknowledges financial support at the Instituto de Astrof\'{\i}sica 
de Andaluc\'{\i}a-{\sc csic} through a {\sc i3p} contract ({\sc i3p-pc2001-1}) funded by the 
European Social Fund.
 
\clearpage

\begin{thebibliography}{}
\bibitem[1999]{bstars}
Aerts, C., De Cat, P., Peeters, E., Decin, L., De Ridder, J.,
Kolenberg, K., Meeus, G., Van Winckel, H., Cuypers, J., \& Waelkens, C., 1999, A\&A, 343, 872
\bibitem[2004]{pms}
Amado, P.J., Moya, A., Su\'arez, J.C., Mart\I n-Ruiz, S., Garrido, R., Rodr\I guez, E.,
Catala, C., \& Goupil, M.J., 2004, MNRAS, 352, L11
\bibitem[2001]{cordoba}
Baglin, A., Auvergne, M., Catala, C., et al. 2002,
in First Eddington Workshop, Cordoba 11-15 June 2001, ed.
J. Christensen-Dalsgaard, \& I. Roxburgh, ESA-SP, 485, 17 
\bibitem[2002]{bru}
Bruntt, H., Catala, C., Garrido, R., Rodr\I guez, E., St\"{u}tz, C., et al., 2002, A\&A, 389, 345
\bibitem[2002]{gdor}
Handler, H., \& Shobbrook, R.R., 2004, MNRAS, 333, 251
\bibitem[1978]{hm}
Hauck, B., \& Mermilliod, M., 1998, A\&AS, 129, 431
\bibitem[2004]{gdor}
Kaye, A.B., Warner, P.B., Guzik, J.A., 2004, in Variable Stars in the Local Group, ed.
D.~W.~Kurtz, \& K.R.~Pollard, ASP Conf. Ser., 310, 474
\bibitem[2001]{mons}
Kupka, F., \& Bruntt, H., 2001, First {\sc corot/mons/most} Ground-based Support Workshop, p.~39,
ed.~C.~Sterken, University of Brussels
\bibitem[2004]{ohp}
Mathias, P., Le Contel, J.-M., Chapellier, E., Jankov, S., Sareyan, J.-P., et al., 2004,
A\&A, 417, 189
\bibitem[1997]{cesam}
Morel, P., 1997, A\&AS, 124, 597
\bibitem[2004]{bb}
Nordstr\"om, B., Mayor, M., Andersen, J., Holmberg, J., Pont, F., J\o rgensen, B.R., Olsen,
E.H., Udry, S., \& Mowlavi, N., 2004, A\&A, 418, 989
\bibitem[1993]{olsen}
Olsen, E.H. 1993, A\&AS, 102, 89
\bibitem[2003]{paperuno}
Poretti, E., Garrido, R., Amado, P.J., Uytterhoeven, K., Handler, G.,
et al., 2003, A\&A, 406, 203 (Paper~I)
\bibitem[1995]{roger}
Rogers, N.Y., 1995, Comm. in Asteroseismology, vol.~78, p.~1
\bibitem[2004]{gaudi}
Solano, E., Catala, C., Garrido, R., Poretti, E., Janot-Pacheco, E., et al.,
2005, AJ, 129, 547 
\bibitem[2004]{rota}
Su\'arez, J.C., Goupil, M.J., Michel, E., Dziembowski, W.A., 
Lebreton, Y., Morel, P., 2004, in Second Eddington Workshop,
Palermo 9--11 April 2003, ed. F.~Favata, S.~Aigrain, \& A.~Wilson,
ESA-SP, 538, 431
\bibitem[2003]{bc}
VandenBerg, D.A., \& Clem, J.L., 2003, AJ, 126, 778
\bibitem[1971]{vani}
Vani\^cek, P. 1971, Ap\&SS 12, 10
\end{thebibliography}
\end{document}